\documentclass[pre,floatfix,twocolumn,showpacs]{revtex4}

\usepackage{amsmath}
\usepackage{amssymb}
\usepackage{graphicx}

\newcommand{\eg}{\emph{e.}$\,$\emph{g.}}
\newcommand{\ie}{\emph{i.}$\,$\emph{e.}}
\newcommand{\etal}{\emph{et}$\,$\emph{al.}}

\newcommand{\un}[1]{\,\text{#1}}

\newcommand{\romad}{{\operatorname{ad}}}
\newcommand{\rombarrier}{{\operatorname{barrier}}}
\newcommand{\rombend}{{\operatorname{bend}}}
\newcommand{\romfree}{{\operatorname{free}}}
\newcommand{\romten}{{\operatorname{ten}}}

\newcommand{\romB}{{\operatorname{B}}}

\newcommand{\romc}{{\operatorname{c}}}
\newcommand{\romd}{{\operatorname{d}}}
\newcommand{\rome}{{\operatorname{e}}}

\newcommand{\tE}{\tilde{E}}
\newcommand{\tw}{\tilde{w}}

\newcommand{\ts}{\tilde{\sigma}}

\newcommand{\VECr}{{\boldsymbol{r}}}

\newcommand{\CALO}{{\mathcal O}}
\newcommand{\CALW}{{\mathcal W}}

\begin{document}

\title{Elastic deformation of a fluid membrane upon colloid binding}

\author{Markus Deserno}
\affiliation{Department of Chemistry and Biochemistry, UCLA, 405 Hilgard Ave, Los Angeles, CA 90095-1569, USA}
\altaffiliation[on leave to: ]{Max-Planck-Institut f\"ur Polymerforschung, Acker\-mann\-weg 10, 55128 Mainz, Germany}
\date{March 28, 2003}
\begin{abstract}
  When a colloidal particle adheres to a fluid membrane, it induces
  elastic deformations in the membrane which oppose its own binding.
  The structural and energetic aspects of this balance are
  theoretically studied within the framework of a Helfrich
  Hamiltonian.  Based on the full nonlinear shape equations for the
  membrane profile, a line of continuous binding transitions and a
  second line of discontinuous envelopment transitions are found,
  which meet at an unusual triple point.  The regime of low tension is
  studied analytically using a small gradient expansion, while in the
  limit of large tension scaling arguments are derived which quantify
  the asymptotic behavior of phase boundary, degree of wrapping, and
  energy barrier.  The maturation of animal viruses by budding is
  discussed as a biological example of such colloid-membrane
  interaction events.
\end{abstract}

\pacs{87.16.Dg, 87.15.Kg, 46.70.Hg}

\maketitle

%%%%%%%%%%%%%%%%%%%%%%%%%%%%%%%%%%%%%%%%%%%%%%%%%%%%%%%%%%%%%%%%%%%%%%%%%%%
%%%%%%%%%%%%%%%%%%%%%%%%%%%%%%%%%%%%%%%%%%%%%%%%%%%%%%%%%%%%%%%%%%%%%%%%%%%

\section{Introduction}\label{sec:intro}

Arguably the most important structural component of all living cells
is the phospholipid bilayer.  It combines the two diametrical tasks of
\emph{partitioning} --  thereby organizing the complex hierarchy
of intracellular biochemical environments -- while at the same time
providing controlled \emph{transport} mechanisms between neighboring
compartments \cite{Lodish}.  The size of particles being transported
spans several orders of magnitude, ranging all the way from
sub-nanometer ions, whose passage through the bilayer is facilitated
by protein channels, up to micron-sized objects engulfed by the large
scale membrane deformations occurring during phagocytosis.

Since cell survival depends on a meticulous balance of these
processes, they are actively controlled and require metabolic energy
to proceed.  Still, there are cases where they happen passively as a
result of generic physical interactions, for instance, a sufficiently
strong adhesion between the particle about to be transported and the
membrane.  An example that has been studied in extensive detail is the
route along which many animal viruses leave their host cell
\cite{GaHe98}.  After entering the cell (typically via receptor
mediated endocytosis or other active processes \cite{SiWh02}) and
completion of the viral replication steps, the new virions have to get
out again.  Many virus families accomplish this by the plasma membrane
wrapping their nucleoprotein capsid and pinching off (``budding'')---a
step which not only sets them free but by which they also acquire
their final coating.  A particularly clean model case is provided by
Semliki Forest Virus, in which case the binding between capsid and
membrane is promoted by viral (``spike''-) proteins
\cite{GaSi74,SiGaHe82,LuKi00}.

A different realization of such a wrapping event is presumably
exploited by an efficient gene transfection system proposed a few
years ago by Boussif \etal\ \cite{BoLe95}.  There, DNA is complexed by
the cationic branched polymer polyethylenimine into a globular
complex, which then enters the cell (as deduced from reporter gene
expression).  Rather than particular targeting signals, the slight net
positive charge of the complex is believed to trigger adhesion and
membrane penetration via an electrostatic interaction with negatively
charged regions of the plasma membrane.

Finally, a great deal of biophysical experimental techniques involve
the attachment of microbeads to membranes.  For instance, one way to
measure cellular tensions involves pulling a thin tether with an
optical tweezer which grabs a bead adhering to the membrane
\cite{HoSh96}.  Cell membranes and subcellular organelles are
routinely probed with an atomic force microscope \cite{HeOb00}.  And a
classical experiment on surface dynamics and locomotion of cells
involves monitoring the centripetal motion of a surface adherent
particle \cite{AbHe70}.  In all cases the object adhering to the cell
will locally deform its plasma membrane, which can be crucial for
interpreting the experimental results.  For example, in the case of
centripetal bead motion it has been noted that adhering particles may
actually become
\emph{engulfed} by the cell without involvement of the endocytosis
machinery, if only the membrane tension is low enough \cite{CaYe01}.

Wrapping and budding processes occur very frequently in cells, but
even though they are extensively studied experimentally, the high
complexity of the real biological situation renders a clear extraction
of underlying physical principles very difficult; it is not even
obvious or undisputed whether an explanation in terms of such
principles is possible.  A better look at the physics is therefore
provided by more easily controllable experiments on the adsorption of
colloids onto model lipid bilayers.  The degree of wrapping of a
colloid by a giant phospholipid vesicle has for instance been studied
by Dietrich \etal\ \cite{DiAn97}.  These authors showed that it can be
quantitatively understood in terms of a balance between adhesion and
elastic energy.  Colloids will change the shape of the vesicle they
adhere to, and this may give rise to attractive interactions between
them \cite{KoRa99}.  If the membrane tension is low enough or the
colloid sufficiently small, bending of the membrane will become an
important contribution which can be strong enough to completely
suppress adhesion \cite{LiDo98,DiWo98,DeGe02}.  The bending stiffness
will also prevent any ``kink'' in the membrane profile at the line of
contact, such that the notion of a contact angle only remains
meaningful in an asymptotic sense and is replaced by the concept of
contact \emph{curvature} \cite{LaLi86,SeLi90,LiSe91,Sei95}.  For
vesicles adhering on flat surfaces, this can be measured by
visualizing the contact zone via reflection interference contrast
microscopy \cite{RaFe95}, but for adhesion of membranes on strongly
curved substrates -- in particular, small colloids -- this option is
not available.  Here, computer simulations can provide means for a
close-up study \cite{NoTa02}, but to date it is still difficult to
perform a quantitative analysis relating the observed
\emph{geometry} to the underlying mesoscopic \emph{elastic}
properties.

Instead of directly measuring the local deformations, one can study
their indirect consequences and from there attempt to deduce some of
their properties.  This of course requires theoretical modeling for
bridging the gap.  For instance, lateral membrane tensions can be
inferred from the force required to pull a thin tether from the
membrane \cite{HoSh96}.  Recent theoretical work by Der\'enyi
\etal\ shows that the force as a function of tether length is quite
subtle and shows an initial oscillatory structure before exponentially
settling down to the asymptotic value \cite{DeJu02}.  Very useful
information might be obtained from this, for example an independent
estimate of the tether \emph{width}.  For less strongly bound beads
Boulbitch has related the unbinding force to the adhesion energy,
membrane bending stiffness and the elasticity of an underlying elastic
network (like the cytoskeleton) by describing the associated membrane
deformation within a small gradient approximation \cite{Bou02}.  And
for the case of colloids adhering to quasispherical vesicles a simple
ansatz for the membrane shape has been shown to yield a structural
phase diagram indicating when the colloid is free, partially wrapped
and fully enveloped \cite{DeGe02}.

This paper extends the work of Ref.~\cite{DeBi} in developing a
detailed theory of the local wrapping behavior in the case of a
constant prescribed lateral membrane tension.  This ``ensemble'' is
particularly relevant for the biological situation, since almost all
cells constantly adjust the amount of lipids in their plasma membrane
in order to maintain its lateral tension at some specific set point
\cite{MoHo01}.  The situation to be discussed here is similar to the one
treated by Boulbitch~\cite{Bou02}, but it will neither be restricted
to small membrane deformations nor to a two-dimensional modeling of
the geometry (which would be appropriate for long cylindrical
colloids).  After setting up the problem and identifying the relevant
energies in Sec.~\ref{sec:gec}, the full nonlinear shape equations of
the membrane profile are studied in Sec.~\ref{sec:full_solution},
leading to the structural wrapping phase diagram.  Sec.~\ref{sec:sge}
then treats the small gradient expansion (\ie, the \emph{linearized}
theory), identifies its range of validity as the regime of low
membrane tension, and derives various asymptotically exact results.
For the opposite regime of large tension several scaling predictions
are deduced in Sec.~\ref{sec:scaling} and validated against the
numerical results from Sec.~\ref{sec:full_solution}.  In the final
Section~\ref{sec:bioexample} the results obtained throughout the paper
are discussed in the framework of the biological example of virus
budding.

%%%%%%%%%%%%%%%%%%%%%%%%%%%%%%%%%%%%%%%%%%%%%%%%%%%%%%%%%%%%%%%%%%%%%%%%%%%
%%%%%%%%%%%%%%%%%%%%%%%%%%%%%%%%%%%%%%%%%%%%%%%%%%%%%%%%%%%%%%%%%%%%%%%%%%%

\section{General energy considerations}\label{sec:gec}

\begin{figure}
  \includegraphics[scale=0.78]{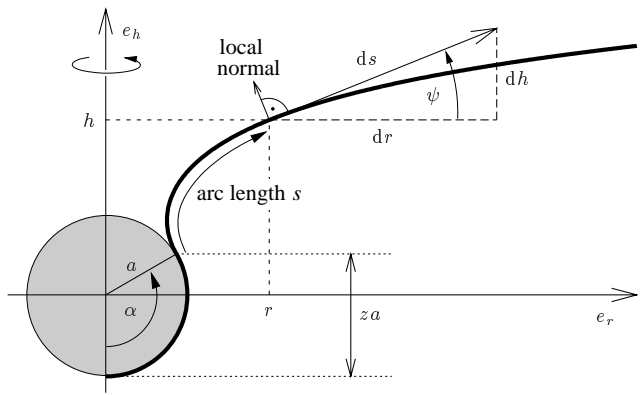}
  \caption{Illustration of the wrapping geometry and membrane
  parameterization.  A membrane adheres partially to a spherical
  colloid of radius $a$ with a degree of wrapping given by
  $z=1-\cos\alpha$.  Cylindrical symmetry around the $e_h$ axis is
  assumed throughout the paper.  The possibility of ``overhangs''
  requires going beyond a Monge-like parameterization, which would
  gives the height $h$ as a function of radial distance $r$.  The
  choice followed in this work is to specify the angle $\psi$ with
  respect to the horizontal as a function of arc-length $s$.}
\label{fig:definitions}
\end{figure}

The aim of this paper is to describe the local deformations of a flat
fluid membrane upon adsorption of a spherical colloid, as depicted
schematically in Fig.~\ref{fig:definitions}.  It will be assumed that
this process can be understood as a balance of the following three
energy contributions: ($i$) Adhesion is driven by a
\emph{contact energy} per unit area, $w$, and opposed by ($ii$) the
requirement to \emph{bend} the membrane as well as ($iii$) the work of
pulling excess membrane toward the wrapping site against a prescribed
\emph{lateral tension}, $\sigma$.  Since the description will
not aim at a microscopic understanding, continuum elasticity theory is
taken as a basis; in particular, the bending energy per unit area will
be described using the standard Helfrich expression \cite{Hel73}
\begin{equation}
  e_\rombend = \frac{1}{2}\kappa(c_1 + c_2 - c_0)^2 + \bar\kappa \,
  c_1c_2 \ , \label{eq:Helfrich}
\end{equation}
where $c_1$ and $c_2$ are the local principal curvatures of the
two-dimensional membrane surface \cite{Krey91}, $c_0$ is the
spontaneous curvature of the membrane and $\kappa$ and $\bar\kappa$
are elastic constants (with units of energy).  In the following a
symmetric membrane is assumed, \ie\ $c_0=0$; and since no topological
changes will be considered, the second term in
Eqn.~(\ref{eq:Helfrich}) can be dropped
\cite{GaussBonnet}.  The tension energy is given by the lateral tension,
$\sigma$, times the excess area.  Note that from tension and bending
constant one can construct a length $\lambda$ -- specific to the
membrane -- according to
\begin{equation}
  \lambda := \sqrt{\frac{\kappa}{\sigma}} \ .
  \label{eq:lambda}
\end{equation}
Membrane deformations on a length scale smaller than $\lambda$
predominantly cost bending energy, while deformations on a larger
scale pay mostly in tension.  An interesting situation occurs when the
two intrinsic lengths of the problem -- the colloid radius $a$ and the
membrane length $\lambda$ -- are of the same order, because then
bending and tension contributions are comparable.  In fact, for a
typical cellular membrane tension of $\sigma \simeq 0.02\un{dyn/cm}$
\cite{MoHo01} and a typical bending modulus of $\kappa \simeq
20\,k_\romB T$ (where $k_\romB T$ is the thermal energy) one obtains
$\lambda \simeq 64\un{nm}$.  Viral capsids are about this big
\cite{GaHe98}, therefore this biological situation sits squarely in
the crossover regime.

Assuming that the adhering membrane remains in its fluid state, its
energy can be calculated easily once the degree of wrapping,
$z=1-\cos\alpha$, is specified, because the shape of the membrane is
known.  The area of the colloid covered by membrane is given by
$A_\romad = 2\pi a^2 z$, which gives an adhesion energy of $E_\romad =
-wA_\romad = -2\pi a^2 z w$.  Using Eqn.~(\ref{eq:Helfrich}), the
bending energy is found to be $E_\rombend = \frac{1}{2} \kappa
(1/a+1/a)^2 A_\romad = 4\pi z\kappa$.  Finally, the work done against
a lateral tension $\sigma$ is proportional through $\sigma$ to the
\emph{excess} area pulled toward the wrapping site, which is $\Delta
A_\romad = \pi a^2z^2$, giving the tension energy $E_\romten = \pi
a^2z^2\sigma$.

It turns out to be advantageous to measure energies in units of the
bending constant $\kappa$ and lengths in units of the colloid radius
$a$.  This suggests the definition of the following three
dimensionless variables:
\begin{equation}
  \tE := \frac{E}{\pi\kappa}
  \quad,\quad
  \tw := \frac{2wa^2}{\kappa}
  \quad,\; \text{and} \quad
  \ts := \frac{\sigma a^2}{\kappa} \ ,
\end{equation}
where numerical factors of $\pi$ and $2$ have been introduced for
later convenience.  Note that the crossover relation $\lambda\simeq a$
corresponds to $\ts\simeq 1$.  In terms of these reduced variables the
total energy of the colloid-membrane-complex is given by
\begin{equation}
  \tE = -(\tw-4)z + \ts z^2 + \tE_\romfree \ ,
  \label{eq:general_energy}
\end{equation}
where $\tE_\romfree = E_\romfree/\pi\kappa$ is the (dimensionless)
energy of the \emph{free} part of the membrane.  This term is not so
easy to calculate, since the membrane shape is not known \emph{a
priori}.  Even though the main purpose of this paper is to determine
$E_\romfree$ and understand its implications on the wrapping process,
it still proves instructive to have a brief look at the problem while
\emph{ignoring} $E_\romfree$.  Minimizing
Eqn.~(\ref{eq:general_energy}) with respect to $z$ (and noting that
the minimum may also be located at the boundaries $z=0$ or $z=2$) it
is found that colloids do not adhere if $\tw<4$, since they cannot pay
the bending price.  Once $\tw>4$, they start to adhere by first being
partially wrapped.  Full envelopment occurs only if $\tw > 4+4\ts$.
In between, the degree of partial wrapping is $z = (\tw-4)/2\ts$, and
both transitions from free to partially wrapped as well as from
partially wrapped to fully enveloped are continuous.

There is one case in which the simplification $E_\romfree=0$ in fact
holds rigorously, and that is the case of zero tension, $\sigma=0$
\cite{LiDo98,DiWo98}.  This can be seen as follows: Consider the
two-parameter family of cylindrically symmetric surfaces $r(h) = a_1
\, \cosh[(h/a-a_2)/a_1]$, which are called \emph{catenoids}.  It is
easily verified that the particular choice $a_1 = z(2-z)$ and $a_2 =
z-1 + \frac{z(2-z)}{2} \ln\frac{z}{2-z}$ implies that this surface
smoothly touches the colloid with a degree of wrapping equal to $z$.
The important point is that such catenoids are \emph{minimal surfaces}
which have zero mean curvature at every point \cite{Krey91}, and hence
no bending energy.  Since ($i$) for $\sigma=0$ the only possible
energy of the free membrane comes from bending, and ($ii$) the bending
energy is positive definite, the catenoid is in fact the minimum
energy shape.  The above simplified description of the problem is thus
exact in the case $\sigma=0$, \ie, for $\tw<4$ the colloid is not
wrapped, and for $\tw>4$ it is completely wrapped, with no energy
barrier impeding the transition \cite{Lipowsky_barrier}.

%%%%%%%%%%%%%%%%%%%%%%%%%%%%%%%%%%%%%%%%%%%%%%%%%%%%%%%%%%%%%%%%%%%%%%%%%%%
%%%%%%%%%%%%%%%%%%%%%%%%%%%%%%%%%%%%%%%%%%%%%%%%%%%%%%%%%%%%%%%%%%%%%%%%%%%

\section{The full solution of the shape profile}\label{sec:full_solution}

By setting $E_\romfree=0$ one neglects in particular any potential
effect that the strongly curved part of the membrane close to the line
of contact may have on the wrapping behavior.  As a next step it is
therefore tempting to approximate $\tE_\romfree$ by a phenomenological
line energy \cite{Lipowsky_barrier}.  However, neither the relation
between the line tension constant and the membrane properties $\kappa$
and $\sigma$ would be known, nor is the implied dependency on the
degree of wrapping -- namely $\sim\sin\alpha=\sqrt{z(2-z)}$ --
supported by more careful studies.  (On a scaling level and in the
high tension limit this question is revisited in
Sec.~\ref{sec:scaling}.)

In order to draw sound conclusions, an explicit treatment of the
membrane deformation is needed.  Three approaches are discussed in the
following: In the present section the equilibrium membrane shape is
determined by minimizing its energy.  Since this leads to complicated
nonlinear differential equations, one largely has to rely on numerical
solutions here.  In the second approach, discussed in
Sec.~\ref{sec:sge}, a restriction to small membrane deformations
renders these differential equations \emph{linear}, and they can be
solved exactly then.  However, the latter approach per construction is
limited to fairly small degrees of wrapping.  Hence, its range of
validity is not obvious and has to be checked against the nonlinear
results.  Finally, in Sec.~\ref{sec:scaling} a scaling analysis is
used to shed light onto the high tension regime.

%%%%%%%%%%%%%%%%%%%%%%%%%%%%%%%%%%%%%%%%%%%%%%%%%%%%%%%%%%%%%%%%%%%%%%%%%%%

\subsection{Energy functional and shape equations}

The energy of the free membrane is the surface integral over the local
bending and tension contributions and is thus a functional of the
shape.  Using the cylindrically symmetric angle--arc-length
parameterization from Fig.~\ref{fig:definitions}, the two principal
curvatures are found to be $(\sin\psi)/r$ and $\dot\psi$, where the
dot indicates a derivative with respect to the arc-length $s$.  The
energy functional can then be written as \cite{JuSe94}
\begin{equation}
  \tE_\romfree = \int_0^\infty \romd s \; L(\psi,\dot\psi,r,\dot
  r,\dot h,\lambda_r,\lambda_h) \ ,
  \label{eq:E_functional}
\end{equation}
where the Lagrange function $L$ is defined by
\begin{eqnarray}
  L & = &
  r\left\{\left(\dot\psi + \frac{\sin\psi}{r}\right)^2 + \frac{2\ts}{a^2}(1-\cos\psi)\right\}
  \nonumber \\
  & & + \; \lambda_r(\dot r-\cos\psi) + \lambda_h(\dot h - \sin\psi) \ .
  \label{eq:Lagrangian}
\end{eqnarray}
The expression in curly brackets contains the bending and tension
contributions, while the two additional terms enforce the nonholonomic
parameterization constraints $\dot r = \cos\psi$ and $\dot h =
\sin\psi$ by means of the Lagrange parameter \emph{functions}
$\lambda_r(s)$ and $\lambda_h(s)$.  Measuring all lengths in units of
$a$, it can be verified that $\tE_\romfree$ depends parametrically on
$\ts$ (and through the boundary conditions on $z$) \emph{but on
nothing else}.  This is important, because it ensures that $\tw$ and
$\ts$ remain the relevant axes for a structural phase diagram and no
new independent variable is introduced by the free part of the
membrane.  It also implies that at given $\ts$ and $z$ the membrane
shape scales with the colloid size $a$.

The Lagrangian $L$ is independent of the arc-length $s$, therefore the
corresponding Hamiltonian is conserved.  Moreover, numerically one
usually integrates systems of first order differential equations.
Both observations suggest to switch to a Hamiltonian description.
After defining the conjugate momenta
\begin{subequations}
\begin{alignat}{2}
  p_\psi & \; = \; \frac{\partial L}{\partial\dot\psi} & \; = \; & 2r\Big(\dot\psi+\frac{\sin\psi}{r}\Big) \ , \label{eq:ppsi} \\
  p_r    & \; = \; \frac{\partial L}{\partial\dot r}   & \; = \; & \lambda_r \ , \;\;\text{and}\\
  p_h    & \; = \; \frac{\partial L}{\partial\dot h}   & \; = \; & \lambda_h \ ,
\end{alignat}
\end{subequations}
a Legendre transform yields
\begin{eqnarray}
  H & = & \dot\psi p_\psi + \dot r p_r + \dot h p_h - L
  \nonumber \\
  & = &
  \frac{p_\psi^2}{4r}-\frac{p_\psi\sin\psi}{r} - \frac{2\ts r}{a^2}(1-\cos\psi)
  \nonumber \\
  & &
  + \; p_r \cos\psi + p_h\sin\psi \ .
  \label{eq:Hamiltonian}
\end{eqnarray}
The shape equations are the associated Hamilton equations:
\begin{subequations}
\label{eq:Hameqs}
\begin{eqnarray}
  \dot\psi & = & \frac{p_\psi}{2r} - \frac{\sin\psi}{r} \ ,
  \\
  \dot r & = & \cos\psi \ ,
  \\
  \dot h & = & \sin\psi \ ,
  \\
  \dot p_\psi & = & \Big(\frac{p_\psi}{r}-p_h\Big)\cos\psi + \Big(\frac{2\ts r}{a^2}+p_r\Big)\sin\psi \ ,
  \\
  \dot p_r & = & \frac{p_\psi}{r}\Big(\frac{p_\psi}{4r} - \frac{\sin\psi}{r}\Big) + \frac{2\ts}{a^2}(1-\cos\psi) \ , \;\;\text{and}
  \\
  \dot p_h & = & 0 \ .
  \label{eq:Ham_ph}
\end{eqnarray}
\end{subequations}
Shape equations of this kind have been studied extensively in the
past, leading, among many other things, to a very detailed
understanding of vesicle conformations (in which case one also needs
to fix surface and volume by additional Lagrange multipliers).  For a
detailed review on this subject see Ref.~\cite{Sei97}.

%%%%%%%%%%%%%%%%%%%%%%%%%%%%%%%%%%%%%%%%%%%%%%%%%%%%%%%%%%%%%%%%%%%%%%%%%%%

\subsection{Boundary conditions}\label{ssec:bc}

The situation to be studied is a colloid wrapped by an initially flat
membrane.  The boundary conditions thus have to ensure that the
membrane touches the colloid smoothly and becomes asymptotically flat
at large radial distances.  At contact, $s=0$, the following must
evidently hold (see again Fig.~\ref{fig:definitions}):
\begin{subequations}
\begin{eqnarray}
  r(0)    & = & a\,\sin\alpha \ , \\
  h(0)    & = & -a\,\cos\alpha \ , \;\; \text{and}\\
  \psi(0) & = & \alpha \ .
\end{eqnarray}
\end{subequations}
The notion of asymptotic flatness can be enforced by requiring the
angle $\psi(s)$ and all of its derivatives to vanish in the limit
$s\rightarrow\infty$ \cite{catenoid_diverges}.  However, it suffices
to demand this for the angle $\psi$ and the meridinal curvature
$\dot\psi$,
\begin{subequations}
\begin{eqnarray}
  \lim_{s\rightarrow\infty} \psi(s) & = & 0 \;\; \text{and}\\
  \lim_{s\rightarrow\infty} \dot\psi(s) & = & 0 \ ,
\end{eqnarray}
\end{subequations}
which implies that curvature and tension energy density vanish if one
moves away from the site where the membrane shape is perturbed by the
adhering colloid.  If $\psi(s)$ vanishes sufficiently rapidly (as it
does for $\sigma>0$, see Sec.~\ref{sec:sge}), all contributions beyond
some large distance $S$ in arc-length will be negligible.  A
convenient way to exploit this is the following: Choose an upper
arc-length $S$ and impose the zero angle condition there.  Hence,
variations of $S$ and $\psi(S)$ are not permitted during functional
minimization, but $r(S)$ and $h(S)$ are still free.  This implies the
additional boundary conditions \cite{JuSe94,CoHi53}
\begin{subequations}
\begin{eqnarray}
  0 & = & \frac{\partial L}{\partial\dot r}\bigg|_{s=S} \! = \; p_r(S) \;\; \text{and}
  \label{eq:prS} \\
  0 & = & \frac{\partial L}{\partial\dot h}\bigg|_{s=S} \! = \; p_h(S) \ .
  \label{eq:phS}
\end{eqnarray}
\end{subequations}
The Hamilton equation (\ref{eq:Ham_ph}) shows that $p_h$ is an
integral of ``motion'', and the boundary condition (\ref{eq:phS})
fixes its value to zero.  Hence, $p_h$ drops out of the problem
everywhere.  The condition on $p_r$ can be made very useful by a
little more thought.  Since for $\ts>0$ the angle $\psi$ converges to
zero in an essentially exponential way (see Sec.~\ref{sec:sge}), the
expression for the Hamiltonian will converge toward $p_r$.  Thus, the
requirement of a flat profile implies $H=H(S)\rightarrow p_r(S)=0$, or
in other words, the profile is flat if the Hamiltonian vanishes.
Using Eqns.~(\ref{eq:ppsi}) and (\ref{eq:Hamiltonian}), this can be
turned into a condition for $p_r$ at the \emph{contact} boundary:
\begin{equation}
  a\,p_r(0) =
  \frac{\sqrt{z(2-z)}}{1-z} \Big\{1 + 2\ts z - \big(a\,\dot\psi_0\big)^2\Big\} \ .
  \label{eq:pr0}
\end{equation}

The only remaining variable for which the contact value is not yet
known is $p_\psi$, or alternatively $\dot\psi$.  At this point a bit
of care is required.  It is well known that the balance between
adhesion energy and elastic membrane deformation results in a boundary
condition on the contact curvature \cite{LaLi86}.  For curved
substrates this becomes \cite{SeLi90}
\begin{equation}
  a\,\dot\psi_0
  :=
  a\, \dot\psi(0)
  =
  1-\sqrt{\tw} \ .
  \label{eq:contact_curvature_condition}
\end{equation}
However, it is crucial to understand that this condition only holds
for the final \emph{equilibrium} shape of the complex.  In the present
case the situation is different, because the aim is to calculate the
energy $E_\romfree$ at any \emph{given} value of the penetration $z$.
Therefore the adhesion balance is \emph{restricted} and
Eqn.~(\ref{eq:contact_curvature_condition}) generally will \emph{not}
hold.  Nevertheless, by later imposing $\partial E/\partial z=0$ for
identifying the equilibrium degree of penetration,
Eqn.~(\ref{eq:contact_curvature_condition}) is recovered.  In fact,
this would be one way to derive it.

Rather than Eqn.~(\ref{eq:contact_curvature_condition}), it is the
condition of \emph{asymptotic flatness} that will determine
$\dot\psi_0$.  In practice, this can be done via a shooting method:
For a trial value of $\dot\psi_0$ integrate the profile, find the
arc-length $s_0$ at which the angle vanishes, \ie\ $\psi(s_0)=0$, and
observe at which radial distance $r(s_0)$ this happens.  Now adjust
$\dot\psi_0$ iteratively and search for the value(s) at which $r(s_0)$
diverges.  One thereby finds the contact curvature as a function of
penetration, $\dot\psi_0(z)$, in what amounts to a nonlinear
eigenvalue problem.

\begin{figure}
\includegraphics[scale=0.9]{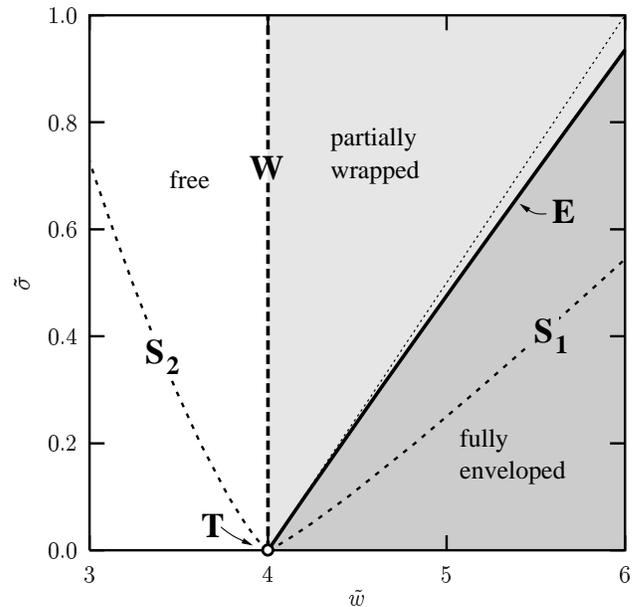}
\caption{Structural wrapping phase diagram in the plane of reduced adhesion
  constant $\tw$ and reduced lateral tension $\ts$, in the low tension
  regime $\ts < 1$ close to the triple point ``T'' $(\tw=4,\ts=0)$.
  The dashed line ``W'' marks the continuous transition at which
  partial wrapping sets in, the bold solid line ``E'' indicates the
  discontinuous transition between partially wrapped and fully
  enveloped, and the short dashed lines ``S$_1$'' and ``S$_2$'' are
  the spinodals belonging to ``E''.  The fine dotted line $\tw=4+2\ts$
  close to ``E'' indicates where the fully wrapped state has zero
  energy.}
\label{fig:pd}
\end{figure}

At this point a side-note seems appropriate.  It turns out that at
$\ts = \ts_\romc = 4.721139\ldots$ the nature of this eigenvalue
problem changes qualitatively, since a region of $z$ values emerges
(at $z_\romc = 1.86289\ldots$) for which there exist \emph{three}
contact curvatures which yield profiles satisfying all boundary
conditions (the function $\dot\psi_0(z)$ develops an S-shape).  The
correct solution has to be identified based on the criterion of lowest
energy.  For somewhat larger values of $\ts$ the lowest energy
solution cannot even be realized physically: The corresponding
curvature $\dot\psi_0$ may become larger than $1/a$, which is
geometrically impossible because the membrane cannot bend \emph{into}
the colloid it adheres to.  This, however, is not a problem in the
present case of an adhesion balance, since the curvature boundary
condition from Eqn.~(\ref{eq:contact_curvature_condition}) implies
that for any wrapping geometry in which the point of detachment is not
fixed the contact curvature has to be \emph{smaller} than $1/a$.  In
fact, it turns out that the accessible range of multivalued contact
curvatures always lies inbetween the transition from partially to
fully wrapped, hence it has no direct consequences on the phase
diagram \cite{bifurcate_relevance}.  But the mathematical properties
of this bifurcation may permit some insight into the general nature of
the solution, which, however, will not be pursued in the present
paper.

%%%%%%%%%%%%%%%%%%%%%%%%%%%%%%%%%%%%%%%%%%%%%%%%%%%%%%%%%%%%%%%%%%%%%%%%%%%

\subsection{Structural phase diagram}\label{ssec:structural_pd}

Numerically performing the calculations indicated above yields the
shape profile and hence, via Eqn.~(\ref{eq:E_functional}), the free
membrane energy for any given value of $\ts$ and $z$.  From
Eqn.~(\ref{eq:general_energy}) one then determines the total energy as
a function of $\ts$, $\tw$, and $z$.  The minimum in $z$ within the
range $[0;2]$ corresponds to the equilibrium state and one obtains the
phase diagram as depicted in Fig.~\ref{fig:pd} \cite{DeBi}.

\begin{figure}
\includegraphics[scale=0.93]{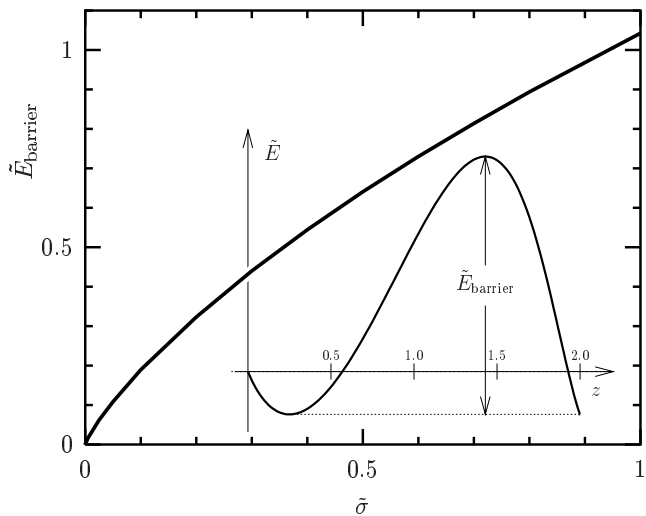}
\caption{Height of the energy barrier $\tE_\rombarrier = E_\rombarrier /
  \pi\kappa$ as a function of reduced tension in the low tension
  regime $\ts<1$, at the value of $\tw$ where the transition from
  partial wrapping to full envelopment occurs.  The inset illustrates
  the shape of the function $\tE(z)$ and defines the concomitant
  energy barrier ($\ts=0.5$ in this example).}
\label{fig:barrier}
\end{figure}

\begin{figure}
\includegraphics[scale=0.88]{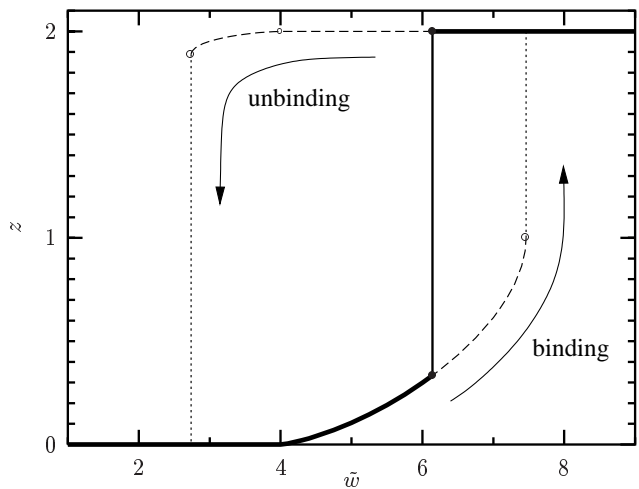}
\caption{Hysteresis loop of crossing the envelopment
  transition by changing the adhesion energy $\tw$ for the particular
  tension $\ts=1$. At $\tw=4$ binding sets in, at $\tw\approx 6.1$ the
  fully enveloped state becomes stable.  However, only at $\tw\approx
  7.5$ does the huge energy barrier of $66\,k_\romB T$ separating it
  from the partially wrapped state vanish.  On the unbinding branch
  the system again remains metastable beyond the actual transition,
  the stable partially wrapped branch of low bindings is entirely
  skipped and ``replaced'' by a metastable partially wrapped branch
  below $\tw=4$, featuring large values of $z$.  At $\tw\approx 2.7$
  the energy barrier for unbinding vanishes.}\label{fig:hysteresis}
\end{figure}

The transition from free to partially wrapped (see dashed line ``W'')
remains unchanged compared to the case where $E_\romfree$ was
neglected.  The reason for this is that -- just like the tension term
-- the energy of the free membrane is of higher than linear order for
small $z$ (this again follows analytically from a small gradient
expansion, see Sec.~\ref{sec:sge}).  The physics is thus determined by
a balance between bending and adhesion alone.  However, the transition
from partial wrapping to full envelopment (the solid line ``E'')
changes significantly: As can be seen in Fig.~\ref{fig:barrier}, an
energy barrier separates the fully and partially wrapped states,
rendering the transition \emph{discontinuous}.  This energy barrier
turns out to be mostly tension (not bending) energy stored in the free
membrane of partially wrapped colloids, and its height can be quite
substantial.  For instance, with $\sigma=0.02\un{dyn/cm}$ (a typical
value for a cellular tension \cite{MoHo01}), $a=30\,$nm (capsid radius
of Semliki Forest Virus, as an example for a colloidal particle to be
wrapped) and $\kappa=20\,k_\romB T$, one finds $\ts
\approx 0.22$ and from that $E_\rombarrier\approx 22\,k_\romB T$.
This barrier is too large to be overcome by thermal fluctuations alone
\cite{barrier_kT}.  However, upon increasing the adhesion energy $\tw$
and thereby going deeper into the region of full envelopment, the
energy barrier separating the partially and the fully wrapped state
decreases, ultimately vanishing at the spinodal line
``S$_1$''. Conversely, once the colloid is fully wrapped, the same
energy barrier prevents the
\emph{un}wrapping transition, and one has to decrease the value of
$\tw$ further in order to remove this barrier---see the second
spinodal ``S$_2$'' in Fig.~\ref{fig:pd}.  Cycling across the
envelopment transition ``E'' thus gives rise to hysteresis, as is
illustrated for the particular case $\ts=1$ in
Fig.~\ref{fig:hysteresis}, for which the energy barrier is
$E_\rombarrier\approx 66\,k_\romB T$ using the same system properties
as above.  Interestingly, this hysteresis is so pronounced that one
``skips'' entirely the partially wrapped region upon unbinding.

\begin{figure}
\includegraphics[scale=0.86]{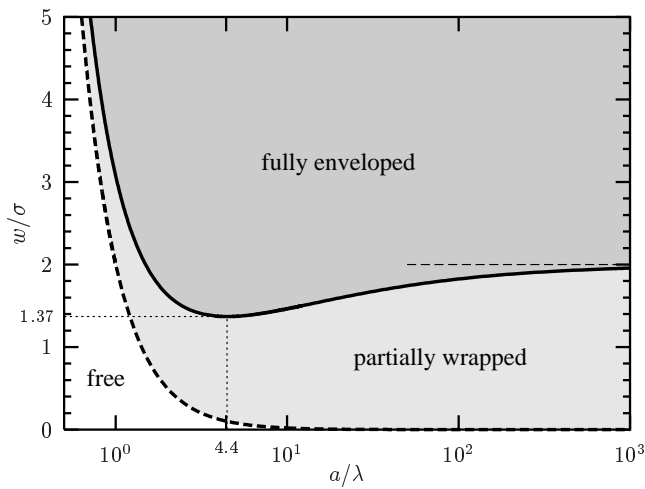}
\caption{Influence of the particle radius $a$ on the wrapping behavior.
  Sufficiently large particles will always at least partially wrap.
  In the range $1.37 \lesssim w/\sigma \le 2$ particles only become
  enveloped if they are neither to small \emph{nor} too large.}\label{fig:a_boundary}
\end{figure}

Both $\tw$ and $\ts$ are proportional to $a^2$; therefore, a scan of
the particle radius $a$ at fixed values of $\kappa$, $\sigma$, and $w$
yields lines in the phase diagram which pass through the origin.  The
shape of the envelopment boundary ``E'' then implies that for
$w/\sigma\lesssim 1.37$ particles will not become fully enveloped,
irrespective of their size, while for $w/\sigma\ge 2$ all sufficiently
large particles are enveloped.  In the small region inbetween, $1.37
\lesssim w/\sigma \le 2$, particles are only enveloped if they are
neither too small \emph{nor} too large.  The asymptotic envelopment
condition for small $a$ coincides with the boundary at which wrapping
sets in, which is $a = \sqrt{2\kappa/w}$ or $w/\sigma =
2\,(\lambda/a)^2$ \cite{LiDo98,DiWo98}.  At the onset of the
possibility of full envelopment, $w/\sigma \simeq 1.37$, the first
particles to be enveloped have a radius $a\simeq 4.4\lambda$.
Fig.~\ref{fig:a_boundary} summarizes these results.

The energy of the free membrane, $E_\romfree$, vanishes not only in
the limit $z \rightarrow 0$ but also in the full-wrapping limit $z
\rightarrow 2$.  This is reminiscent of the case of an ideal neck
connecting two vesicles \cite{FoMi94}.  The reason is essentially that
the membrane shape locally approaches a catenoid
\cite{catenoid_lambda}.  This fact is very convenient, because
knowledge of the exact energy of the enveloped state greatly
simplifies the discussion of the structural transitions.  This will be
further exploited in Sec.~\ref{sec:sge}.  As an immediate consequence,
it becomes possible to estimate the point of envelopment by comparing
the known energy of the fully enveloped state not with the partially
wrapped state but simply with the free state, namely, $E=0$.  This
gives the boundary $\tw = 4+2\ts$, which is also plotted in
Fig.~\ref{fig:pd} and which actually becomes asymptotic to the real
phase boundary in the limit $\ts\rightarrow 0$ (in a complicated
logarithmic fashion).  Note that this line differs from the phase
boundary of the case where $E_\romfree$ had been neglected by a factor
2 in the slope (\ie, the prefactor of $\ts$)---and in a maybe
unexpected way: The region in the phase diagram belonging to fully
enveloped states grows at the expense of partially wrapped states.
Even though bending and tension energy work \emph{against} adhesion,
they can actually \emph{promote} wrapping.  The reason is that
partially wrapped states with a large penetration can \emph{lower}
their energy by completing the wrapping, which provides another means
to understand why the transition is discontinuous
\cite{fusion_speculation}.  The same has been found for colloids
adhering to quasispherical vesicles \cite{DeGe02}.

For increasing $\ts$ the bending energy should ultimately become
negligible compared to the tension.  Indeed, in the limit $\kappa
\rightarrow 0$ the term $E_\romfree$ vanishes, because the
membrane is flat immediately after detaching (smoothness of the slope
is no longer required).  The equilibrium penetration of partially
wrapped colloids, as deduced from Eqn.~(\ref{eq:general_energy}), is
then $z=\tw/2\ts$.  This equation can be rewritten as $w =
\sigma[1+\cos(\pi-\alpha)]$ and is thereby recognized as the
\emph{Young-Dupr\'{e} equation} \cite{RoWi02}, which relates
adhesion and tension to the contact angle, here $\pi-\alpha$. The
envelopment boundary ``E'' consequently occurs at $\tw = 4\ts$, \ie,
where the penetration is $z=2$, or, equivalently, where the contact
angle vanishes and the membrane completely ``wets'' the colloid.

\begin{figure}
\includegraphics[scale=0.86]{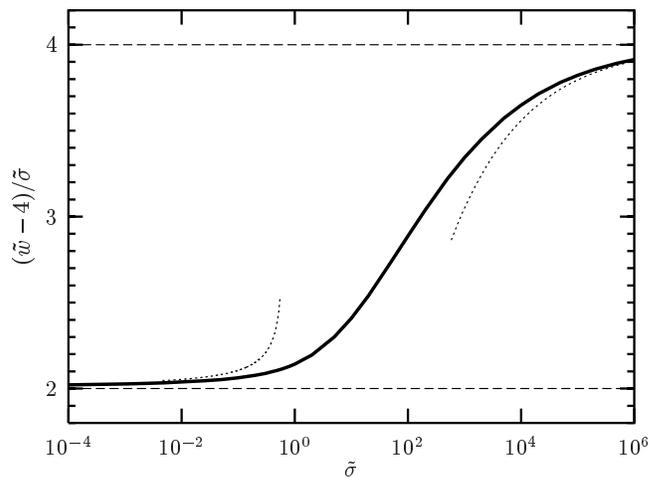}
\caption{Phase boundary between partially and fully wrapped state plotted
  against ten decades of reduced tension, $\ts$.  The combination
  $(\tw-4)/\ts$ is seen to cross over from the value 2 at small $\ts$,
  which follows from the small gradient expansion, to the
  Young-Dupr{\'e}-limit 4, which results when the energy $E_\romfree$
  of the free membrane is negligible.  The two dotted curves at small
  and large $\ts$ correspond to the small gradient estimate
  (\ref{eq:sg_pb_approx}) and the scaling prediction
  (\ref{eq:W_scaling}), respectively.}\label{fig:W_sig_logscan}
\end{figure}

On the basis of these results one expects a cross-over from the small
gradient asymptotic phase boundary $\tw=4+2\ts$, valid below $\ts
\simeq 1$, to a large tension limit $\tw = 4\ts \simeq 4+4\ts$.
Fig.~\ref{fig:W_sig_logscan} confirms this.  However, it is quite
remarkable how many orders of magnitude of variation of the reduced
tension it takes to establish the transition toward the high tension
asymptotic: At $\ts=1$ the curve is about 0.142 away from the zero
tension asymptotic; getting as close as that to the high tension
asymptotics requires $\ts \simeq 2\times 10^5$.  Over the intervening
five orders of magnitude of reduced tension, the influence of bending
and tension cannot be easily disentangled.  In Sec.~\ref{sec:scaling}
it will be shown that the Young-Dupr\'{e}-limit is reached in a power
law fashion with a small exponent $1/3$, which partly explains this
slow crossover.

It may be worth pointing out that the large range of values of $\ts$
is not experimentally unreasonable, because each of the three
variables entering the reduced tension can vary by a few orders of
magnitude: Membrane tensions between $0.01\un{dyn/cm}$ and
$1\un{dyn/cm}$ are typical \cite{MoHo01}, as are bending constants
between $1\,k_\romB T$ and $100\,k_\romB T$ \cite{SeLi95}.  Assuming
colloidal radii between $20\un{nm}$ and $2\un{$\mu$m}$ yields a range
for $\ts$ from about $10^{-2}$ up to $10^6$.

%%%%%%%%%%%%%%%%%%%%%%%%%%%%%%%%%%%%%%%%%%%%%%%%%%%%%%%%%%%%%%%%%%%%%%%%%%%
%%%%%%%%%%%%%%%%%%%%%%%%%%%%%%%%%%%%%%%%%%%%%%%%%%%%%%%%%%%%%%%%%%%%%%%%%%%

\section{Small gradient expansion}\label{sec:sge}

One particular result from the numerical solution of the nonlinear
problem is the following: For sufficiently small tension the
equilibrium penetration shortly before envelopment ensues is quite
small (see the inset in Fig.~\ref{fig:barrier}), as is the concomitant
perturbation of the flat membrane.  Therefore, this region of the
phase diagram should be amenable to an approximate treatment of the
differential equations which corresponds to a lowest order expansion
around the flat profile.

%%%%%%%%%%%%%%%%%%%%%%%%%%%%%%%%%%%%%%%%%%%%%%%%%%%%%%%%%%%%%%%%%%%%%%%%%%%

\subsection{Functional and linear shape equations}

If the shape of the membrane is only weakly perturbed, a Monge
representation giving the profile height $h$ as a function of the
position $\VECr=(x,y)$ in the reference plane is applicable.  Bending
plus tension energy can then be written as
\begin{equation}
  E = \int \romd^2 \boldsymbol{r} \sqrt{1+(\nabla h)^2}
  \left\{ \frac{\kappa}{2}
    \left[ \nabla\cdot
      \frac{\nabla h}{\sqrt{1+(\nabla h)^2}}
    \right]^2 \!\!\! + \sigma
  \right\}\! \ ,
\end{equation}
where $\nabla$ is the two-dimensional nabla operator in the reference
plane.  Expanding the two terms in the integrand up to lowest order in
$\nabla h$ gives the small gradient expansion of the energy functional
\begin{equation}
  E = \int \romd^2 \boldsymbol{r} \left\{ \frac{\kappa}{2}
  \left(\nabla^2 h\right)^2+\frac{\sigma}{2} \left(\nabla h\right)^2 \right\} \ .
  \label{eq:sg_functional}
\end{equation}
The functional variation $\delta E = 0$ finally yields the
\emph{linear} shape equation \cite{next_order}:
\begin{equation}
  \nabla^2 \left(\nabla^2 -\lambda^{-2}\right)h=0 \ ,
  \label{eq:linear_shape_equation}
\end{equation}
where $\lambda$ is the length introduced in Eqn.~(\ref{eq:lambda}).

%%%%%%%%%%%%%%%%%%%%%%%%%%%%%%%%%%%%%%%%%%%%%%%%%%%%%%%%%%%%%%%%%%%%%%%%%%%

\subsection{Equilibrium profile and energy}

\begin{figure}
\includegraphics[scale=1.04]{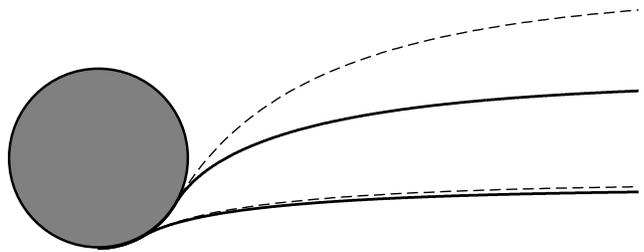}
\caption{Exact membrane profiles (solid curves) and small gradient
  approximation (dashed curves) for two fixed detachment angles
  $\alpha = 30^\circ$ and $\alpha = 60^\circ$.  The reduced tension is
  $\ts = 0.1$.}
\label{fig:profile_comp}
\end{figure}

The differential equation (\ref{eq:linear_shape_equation}) is solved
by eigenfunctions of the Laplacian corresponding to the eigenvalues
$0$ and $\lambda^{-2}$.  In the present cylindrical symmetry the
general solution can therefore be written as
\begin{equation}
  h(r)=h_1+h_2\ln(r/\lambda) + h_3 K_0(r/\lambda) +h_4 I_0(r/\lambda) \ ,
\end{equation}
where $K_0$ and $I_0$ are modified Bessel functions.  Since $I_0(r)$
diverges as $r\rightarrow\infty$, the condition of a flat profile
requires $h_4=0$.  And the coefficient $h_2$ has to vanish since
otherwise the energy density is not integrable for $r \rightarrow
\infty$.  The two remaining coefficients are obtained by fixing height
and slope of the profile at the point where it touches the colloid.  A
straightforward calculation then gives the small gradient profile
\begin{equation}
  \frac{h(r)}{a}
  =
  z - 1 + \frac{\lambda}{a}\frac{k}{1-z}\frac{K_0(ka/\lambda)-K_0(r/\lambda)}{K_1(ka/\lambda)} \ ,
  \label{eq:sg_profile}
\end{equation}
where the abbreviation $k=\sin\alpha=\sqrt{z(2-z)}$ has been used.
Fig.~\ref{fig:profile_comp} gives an example how the small gradient
prediction of the profile compares to the full solution.  If the
detachment angle $\alpha$ is sufficiently small, the overall membrane
deformation remains also small, and the profile from the linearized
theory follows the full solution quite accurately.  However, for a
somewhat larger $\alpha$ significant deviations appear: the membrane
deformation is predicted to be substantially larger than it actually
is.  It is worth pointing out that a good understanding of the profile
is important if one attempts to infer physical properties of the
membrane or the complex by measuring the membrane deformation and
working backwards.  Using the linearized prediction of the profile
then would lead to incorrect conclusions, for instance to an
underestimation of the degree of wrapping.

The energy corresponding to the optimized membrane shape is obtained
by inserting the profile (\ref{eq:sg_profile}) back into the
functional (\ref{eq:sg_functional}).  The necessary integration can be
performed analytically, yielding \cite{DeBi}
\begin{equation}
 \tE_\romfree = \frac{a}{\lambda} \left(\frac{k^3}{1-k^2}\right)
  \frac{K_0\left(ka/\lambda \right)}{K_1\left(ka/\lambda \right)} \ .
  \label{eq:sg_energy}
\end{equation}

%%%%%%%%%%%%%%%%%%%%%%%%%%%%%%%%%%%%%%%%%%%%%%%%%%%%%%%%%%%%%%%%%%%%%%%%%%%

\subsection{Exact asymptotic results}

The small gradient expansion becomes asymptotically exact in the limit
of vanishing penetration, $z\rightarrow 0$.  It is then also
permissible to replace the expression (\ref{eq:sg_energy}) by its
small $z$ expansion
\begin{equation}
  \tE_\romfree
  =
  -2\ts z^2\Big(2\gamma + \ln\frac{\ts z}{2}\Big) + \CALO(z^3) \ ,
  \label{eq:sg_energy_approx}
\end{equation}
where $\gamma = 0.5772\ldots$ is the Euler-Mascheroni constant.  Note
that up to an (important) logarithmic correction this has the form of
a \emph{tension} energy.  Inserting this into
Eqn.~(\ref{eq:general_energy}) one obtains the small gradient free
energy, up to quadratic order in the penetration $z$, and one can
discuss the phase behavior.

As already mentioned in Sec.~\ref{ssec:structural_pd}, the energy of
the free membrane is of higher than linear order for small $z$, hence
the transition from free to partially wrapped is dictated by the
balance between bending and adhesion alone, giving the phase boundary
$\tw=4$.  The equilibrium penetration follows from $\partial\tE /
\partial z = 0$, which can be rewritten as
\begin{equation}
  W\rome^W = -\frac{\tw-4}{8}\rome^{2\gamma}
  \quad\text{with}\quad
  W := 2\gamma + \ln\frac{\ts z}{2} \ .
  \label{eq:W}
\end{equation}
The solution of this transcendental equation is known as the Lambert
$\CALW$ function \cite{CoGo96}, and in the present situation it is the
branch $-1$ which is needed.  One thus obtains
\begin{equation}
  z
  =
  \frac{2\,\rome^{-2\gamma}}{\ts}\rome^W
  =
  \frac{2\,\rome^{-2\gamma}}{\ts W}W\rome^W
  =
  -\frac{\tw-4}{4\ts W}
  \label{eq:z_sg}
\end{equation}
with
\begin{equation}
  W = W(\tw) = \CALW_{-1}\Big(-\frac{\tw-4}{8}\rome^{2\gamma}\Big) \ .
\end{equation}
Equation (\ref{eq:z_sg}) thus gives the penetration as a function of
$\tw$ and $\ts$.  For $x\rightarrow 0^-$ the function $\CALW_{-1}(x)$
diverges to $-\infty$ in a logarithmic way \cite{LambertW}, therefore
the penetration $z(\tw)$ increases at $\tw=4$ for all values of the
tension linearly up to a logarithmic correction.  Note also that its
dependence on the tension is very simple, namely, just inversely
proportional.

The expression (\ref{eq:z_sg}) is of course not valid for all $\tw>4$,
because the system must also cross the transition toward the enveloped
state.  At this point one has to make use of a piece of information
known from the nonlinear studies, namely, the energy of the fully
enveloped state.  It was found that the energy of the free membrane
vanishes as $z$ approaches 2, hence the full envelopment boundary is
given by the simultaneous solution of $\partial\tE/\partial z = 0$ and
the additional equation
\begin{equation}
  \tE(z) = \tE(2) = -2(\tw-4)+4\ts \ .
  \label{eq:E2}
\end{equation}
By eliminating the logarithmic term between those two equations one
obtains a quadratic equation for $z$.  After inserting its solution
into Eqn. (\ref{eq:z_sg}), the final expression can be solved for
$\ts$, and one arrives at the phase boundary
\begin{equation}
  \ts = \frac{\tw-4}{4}\left[1+\sqrt{1+\frac{1}{2W}+\frac{1}{(2W)^2}}\right] \ .
  \label{eq:sg_pb}
\end{equation}
Remembering the divergence of $W$ as $\tw\rightarrow 4^+$,
Eqn.~(\ref{eq:sg_pb}) shows that in the limit of weak binding the
phase boundary approaches $\tw=4+2\ts$, as has been anticipated from
the numerical results in Sec.~\ref{ssec:structural_pd} (see also
Fig.~\ref{fig:pd}).  Using the lowest order approximation
$\CALW_{-1}(x)\sim\ln|x|$ at $x\rightarrow 0^-$ \cite{LambertW} and
expanding the square root in Eqn.~(\ref{eq:sg_pb}), on gets the
approximate asymptotic phase boundary
\begin{equation}
  \ts \simeq
  \frac{\tw-4}{2}\left[1+\frac{1}{8}\Big(2\gamma + \ln\frac{\tw-4}{8}\Big)^{-1}\right] \ ,
  \label{eq:sg_pb_approx}
\end{equation}
which is indicated by the left dotted curve in
Fig.~\ref{fig:W_sig_logscan}.  The expression (\ref{eq:sg_pb}) is
significantly more accurate, but it requires the function $\CALW_{-1}$
to be evaluated.  In any case, since $\CALW_{-1}(x)$ is only real for
$-1/\rome \le x < 0$, even the full expression (\ref{eq:sg_pb}) exists
only up to $\tw = 4+8\,\rome^{-1-2\gamma}\approx 4.928$ or,
equivalently, $\ts =
\rome^{-1-2\gamma}/(2-\sqrt{3}) \approx 0.433$.  At the upper boundary
for $\ts$ one finds $z = 2(2-\sqrt{3}) \approx 0.536$.  Larger
penetrations than this cannot be described within the quadratic
approximation (\ref{eq:sg_energy_approx}) to the small gradient energy
(\ref{eq:sg_energy}).

Finally, the value of the penetration $z$ on the discontinuous phase
boundary can be obtained by eliminating $\tw$ between the two defining
equations $\partial\tE/\partial z = 0$ and (\ref{eq:E2}).  Solving the
remaining equation for $\ts$ yields
\begin{equation}
  \ts =
  \frac{2}{z}\,\exp\left\{-\frac{4-z^2}{2z(4-z)}-2\gamma\right\}
  \stackrel{z \ll 1}\simeq
  \frac{2}{z}\,\exp\Big\{-\frac{1}{2z}\Big\} \ .
  \label{eq:ts_z}
\end{equation}
The second approximate relation can also be solved in terms of the
Lambert $\CALW$ function:
\begin{equation}
  z 
  \stackrel{\ts \ll 1}\simeq
  -\frac{1}{2\,\CALW_{-1}(-\ts/4)}
  \stackrel{\text{\cite{LambertW}}}{\approx}
  -\frac{1}{2}\left[\ln\frac{\ts}{4} - \ln\Big|\ln\frac{\ts}{4}\Big|\right]^{-1} .
\end{equation}

In the limit $\ts\rightarrow 0$, \ie, when approaching the triple
point, the penetration on the discontinuous phase boundary vanishes.
Hence, the jump in order parameter approaches $2$, \ie, the transition
becomes increasingly discontinuous at smaller $\ts$.  However,
Fig.~\ref{fig:barrier} demonstrates that the barrier vanishes in the
limit $\ts\rightarrow 0$, so from this point of view the transition
becomes more continuous.  The triple point $(\tw=4;\ts=0)$ is thus
quite unusual.  Another peculiarity is that along the phase boundary
$z$ does not approach 0 in an algebraic way; rather,
Eqn.~(\ref{eq:ts_z}) shows that $z(\ts)$ has an essential singularity
at $\ts=0$.  All this is related to the fact that the small gradient
expression for the energy is \emph{not} a conventional Landau
expansion in powers of the order parameter, $z$, since the quadratic
term has an additional logarithmic factor.  This lies at the heart of
all logarithmic corrections encountered above (manifest also in the
occurrence of the function $\CALW$), and it renders the standard
classification schemes for critical points inapplicable here.

\vspace*{0.5em}

The exact asymptotic phase boundary can be obtained, because
information about the energy of the fully wrapped state is available.
However, for the \emph{barrier} the situation is different: Even if
the equilibrium penetration is very small, the location of the barrier
(\ie, the penetration $z_\rombarrier$ at which the energy is largest)
occurs at large $z$ (see \eg\ the inset of Fig.~\ref{fig:barrier}).
In fact, numerical evidence suggests that $\lim_{\ts\rightarrow
0}z_\rombarrier=1$ from above.  It is therefore impossible to obtain
the height of the barrier by extending the above small gradient
analysis.

%%%%%%%%%%%%%%%%%%%%%%%%%%%%%%%%%%%%%%%%%%%%%%%%%%%%%%%%%%%%%%%%%%%%%%%%%%%

\section{Scaling laws in the high tension limit}\label{sec:scaling}

As the tension grows, so does the equilibrium penetration on the phase
boundary toward the fully enveloped state.  The location of this
transition can then no longer be obtained within the small gradient
framework of the previous section.  Still, the numerical results in
Sec.~\ref{ssec:structural_pd} strongly suggested that the system
displays a well defined and simple asymptotic behavior in the high
tension limit (see, \eg, Fig.~\ref{fig:W_sig_logscan}).
Unfortunately, treating the curvature as a small perturbation to the
tension is tricky, because this leads to a so-called ``boundary layer
problem'': The solution features a finite variation over a range which
vanishes in the perturbative limit \cite{BenOrs99}.  In the present
case, the membrane has to bend away from the colloid toward the flat
plane (\ie, $\psi(s)$ has to change from $\alpha$ to $0$) in a region
of vanishing arc-length.  Typically, such problems are dealt with by a
subtle matching procedure (an example is provided by the treatment of
an ideal neck in Ref.~\cite{FoMi94}).  Somewhat less ambitious, the
current section shows how the asymptotic behavior can be quantified by
starting with reasonable scaling assumptions about the boundary layer.
Still, the resulting formulas will turn out to be remarkably robust.

A useful observation to start with is that for large $\ts$ the
equilibrium penetration approaches $z=\tw/2\ts$ (see the discussion of
the Young-Dupr\'{e}-limit $\kappa\rightarrow 0$ in
Sec.~\ref{ssec:structural_pd}).  Using the contact curvature boundary
condition (\ref{eq:contact_curvature_condition}), this would predict
the asymptotic relation
\begin{equation}
  a\dot\psi_0
  =
  1-\sqrt{\tw}
  \stackrel{\ts\gg 1}{\sim}
  -\sqrt{2z\ts} \ .
  \label{eq:psidot0_scale}
\end{equation}
This suspicion is indeed confirmed by a check with the numerical
results (data not shown).  The proportionality to $\sqrt{\ts} =
a/\lambda$ is not surprising, since $\lambda$ is the typical length on
which the membrane bends.  However, the proportionality to
$\sqrt{2z}=2\sin\frac{\alpha}{2}$ is not obvious \cite{psidot0}.

Equation (\ref{eq:psidot0_scale}) can be used to infer the asymptotic
behavior of several more variables, by virtue of the following scaling
argument.  Its aim is to estimate the energy of the free part of the
membrane, which for large $\ts$ is largely stored in a small toroidal
rim at contact (this is the boundary layer).  This toroid has the
axial radius $a\sin\alpha$ and a typical meridinal radius which scales
like $1/\dot\psi_0$.  Its area is thus proportional to
$(a\sin\alpha)/\dot\psi_0$, and the tension contribution becomes
\begin{equation}
  E_\romfree^{\text{ten}}
  \sim
  \sigma\times\frac{a\sin\alpha}{\dot\psi_0}
  \sim
  \kappa\sqrt{\ts}\sqrt{2-z}.
  \label{eq:scaling_ten}
\end{equation}
The two principal curvatures are $1/a$ and $\dot\psi_0$, where the
second one clearly dominates in the high tension limit.  Hence, the
bending energy of this torus scales like
\begin{equation}
  E_\romfree^{\text{bend}}
  \sim
 \kappa\times\frac{a\sin\alpha}{\dot\psi_0}\times(\dot\psi_0)^2
  \sim
  \kappa\sqrt{\ts} z\sqrt{2-z}
  \sim
  \kappa\sqrt{\ts} \sqrt{2-z} \ ,
  \label{eq:scaling_bend}
\end{equation}
where in the last step the prefactor $z$ has been dropped, since for
high tension the equilibrium penetration at the transition is close to
2.  Eqns.~(\ref{eq:scaling_ten}) and (\ref{eq:scaling_bend}) show that
in the limit of large tension and close to full wrapping the energy of
the free membrane can be written in the following way
\begin{equation}
  \tE_\romfree
  \simeq
  2 \, A \, \sqrt{\ts} \, \sqrt{2-z}
  \qquad
  (\ts\gg 1, z \approx 2) \ ,
  \label{eq:E_highsigma_scaling}
\end{equation}
where the proportionality factor $A$ does not depend on $\ts$ or $z$,
and the additional factor $2$ is included for convenience.

It is worth pointing out that the $z$--dependence of the scaling form
in Eqn.~(\ref{eq:E_highsigma_scaling}) can also be understood in the
following intuitive way: If one conceives of the strongly curved
region at detachment as giving rise to a \emph{line energy}, it
follows that $E_\romfree$ ought to be proportional to the length of
this line, which is $a\sqrt{z(2-z)}$.  For $z$ close to 2 this has the
same characteristic variation $\sqrt{2-z}$ as
Eqn.~(\ref{eq:E_highsigma_scaling}).  However, it must be noted that
this form holds only in the double limit of large tension and large
penetration.  Generally, $E_\romfree$ is not well represented by a
simple line energy alone.

\begin{figure}
\includegraphics[scale=0.84]{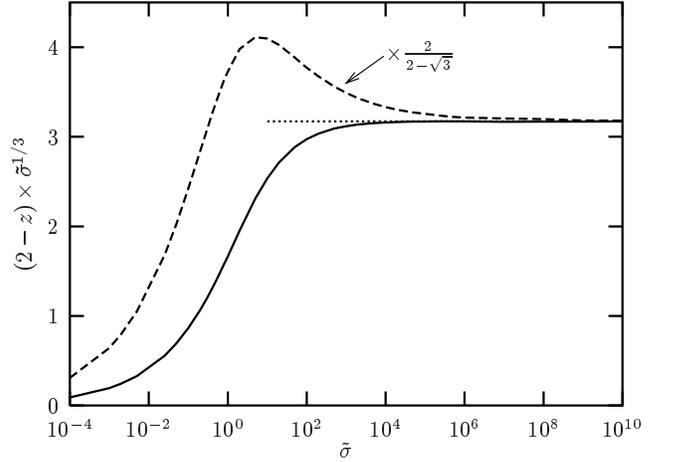}
\caption{Scaling plot for the penetration $z$ on the phase boundary
  (solid) and the location of the barrier, $z_\rombarrier$ (dashed).
  According to Eqns.~(\ref{eq:z_pb_scaling}) and
  (\ref{eq:z_barrier_scaling}) the combination $(2-z)\ts^{1/3}$ should
  approach a constant value, which should be the same for both cases
  if the latter is multiplied by the additional factor
  $2/(2-\sqrt{3})$.  The dotted line indicates the asymptotic limit
  $A^{2/3}\approx 3.17$.}\label{fig:z_scale}
\end{figure}

One can now insert the expression (\ref{eq:E_highsigma_scaling}) into
Eqn.~(\ref{eq:general_energy}) and discuss the phase behavior.
Eliminating $\tw$ between the two equations $\tE(z)=\tE(2)$ and
$\partial\tE(z)/\partial z=0$ gives the penetration $z$ at the
transition as a function of reduced tension:
\begin{equation}
  z 
  \stackrel{\ts\gg 1}{\simeq}
  2 - A^{2/3}\ts^{-1/3} \ .
  \label{eq:z_pb_scaling}
\end{equation}
The high tension limit of the penetration is thus reached in an
algebraic way with an exponent $-1/3$; see Fig.~\ref{fig:z_scale}.
Eliminating $z$ instead of $\tw$ gives the envelopment boundary:
\begin{equation}
  \frac{\tw-4}{\ts}
  \stackrel{\ts\gg 1}{\simeq}
  4 - 3\,A^{2/3}\ts^{-1/3} \ ,
  \label{eq:W_scaling}
\end{equation}
showing that its asymptotic value is also reached algebraically with
an exponent $-1/3$; see Fig.~\ref{fig:W_sig_logscan}.

Information on the barrier can be obtained by further studying the
scaling energy.  After inserting the phase boundary
(\ref{eq:W_scaling}) back into the energy one determines the location
of the maximum via $\partial\tE(z)/\partial z=0$:
\begin{equation}
  z_\rombarrier
  \stackrel{\ts\gg 1}{\simeq}
  2 - \frac{2-\sqrt{3}}{2}A^{2/3}\ts^{-1/3} \ .
  \label{eq:z_barrier_scaling}
\end{equation}
The location of the barrier thus reaches the asymptotic value 2 in the
same way as the location of the transition, only the prefactor is
different by $(2-\sqrt{3})/2$.  Fig.~\ref{fig:z_scale} also shows a
scaling plot of the location of the barrier, in which this additional
factor has been explicitly included.  The fact that both curves in
Fig.~\ref{fig:z_scale} approach the same limit indicates that the
present scaling argument predicts more than the exponent: it correctly
predicts the \emph{ratio of the prefactors} as well.

\begin{figure}
\includegraphics[scale=0.84]{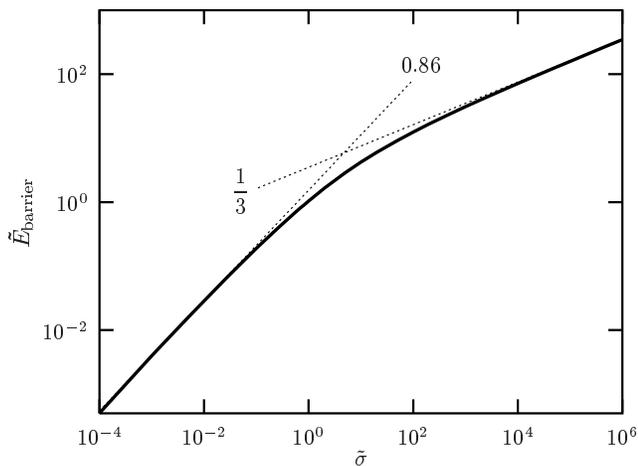}
\caption{Barrier height for the transition between partially wrapped and
  fully enveloped as a function of reduced tension $\ts$ on a double
  logarithmic scale.  The two dotted lines indicate the two different
  regimes: For high tension the barrier height scales with an exponent
  $1/3$, while an empirical power-law fit to the low tension regime
  gives the value $0.86$
  \cite{low_tension_exponent}.}\label{fig:barrier_scale}
\end{figure}

Finally, the barrier height is determined as the difference between
$\tE(z_\rombarrier)$ and $\tE(2)$, for which one finds
\begin{equation}
  \tE_\rombarrier
  \stackrel{\ts\gg 1}{\simeq}
  \frac{3}{4}\,\big(2\sqrt{3}-3\big) \, A^{4/3} \ts ^{1/3} \ .
  \label{eq:E_barrier_scaling}
\end{equation}
This is illustrated in Fig.~\ref{fig:barrier_scale}.  The two
asymptotic power laws meet at the crossover point $\ts_{\text{cross}}
\approx 4.72$.  The numerical value is intriguingly close to the critical
tension $\ts_\romc$ mentioned at the end of Sec.~\ref{ssec:bc}, but
this is probably coincidental.

That the above scaling argument gives the correct relation between the
prefactors can also be checked in the following way: Each of the
equations (\ref{eq:z_pb_scaling}), (\ref{eq:W_scaling}),
(\ref{eq:z_barrier_scaling}), and (\ref{eq:E_barrier_scaling})
describes a scaling relation for a different variable, but the
prefactors all involve $A$.  Numerically one can determine $A$ by an
asymptotic fit to the high tension values of these four variables,
determined from the nonlinear studies of Sec.~\ref{sec:full_solution}.
In all cases one finds the \emph{same} result: $A\approx 5.650$.  The
scatter among the four results relative to the average value is very
small, only about $6\times10^{-4}$.

%%%%%
%%
%% from Eqn  eq:z_pb_scaling:       A  =  5.64927              fit : [1e3:]
%% from Eqn  eq:W_scaling:          A  =  5.64708871196264     fit : [1e6:]
%% from Eqn  eq:z_barrier_scaling:  A  =  5.64875              fit : [1e4:]
%% from Eqn  eq:E_barrier_scaling:  A  =  5.65644815416503     fit : [1e6:]
%%---------------------------------------------
%%               average           <A> =  5.65039
%%  standard deviation <(A-<A>)^2>^0.5 =  0.0035897
%%                ratio                =  0.000635301
%%
%%%%%

%%%%%%%%%%%%%%%%%%%%%%%%%%%%%%%%%%%%%%%%%%%%%%%%%%%%%%%%%%%%%%%%%%%%%%%%%%%
%%%%%%%%%%%%%%%%%%%%%%%%%%%%%%%%%%%%%%%%%%%%%%%%%%%%%%%%%%%%%%%%%%%%%%%%%%%

\section{Discussion of a biological example}\label{sec:bioexample}

In the previous three Sections a theoretical description of the
adhesion and wrapping behavior between a colloid and a fluid membrane
in terms of continuum elasticity theory has been developed, using the
full nonlinear shape equations, their small gradient expansion, and
scaling arguments.  In this final section the results obtained are
used to again make contact with a biological application of such
wrapping events mentioned in the Introduction, namely, the maturation
of animal viruses by budding.

While the reduced tension $\ts$ used throughout the paper can span
many orders of magnitude, it is important to realize that in a
\emph{biological} context the variation is more restricted.
Tensions of cellular membranes reported in the literature vary between
$0.003\un{dyn/cm}$ and $1\un{dyn/cm}$ \cite{MoHo01}.  Larger values
soon result in a structural failure of the bilayer.  On the other
hand, typical bending constants of membranes are in the range of a few
tens up to about a hundred $k_\romB T$.  From these numbers one finds
that the characteristic membrane length $\lambda$ from
Eqn.~(\ref{eq:lambda}) varies roughly between ten and at most a few
hundred nanometers.  Interestingly, this coincides with the range of
particle sizes for which the scenario treated in this paper is
biologically meaningful, for quite different reasons: Particles much
smaller than $10\un{nm}$ are more likely to be transported across a
biomembrane by means of \emph{channels}, while wrapping of particles
much bigger than a few hundred nanometers can no longer be described
without considering the concomitant significant rearrangements of the
cytoskeleton.  Hence, if wrapping events of the kind discussed in this
paper take place on cellular membranes, they are bound to occur in the
regime in which the reduced tension $\ts=(a/\lambda)^2$ is of order 1.

A prominent class of colloidal particles exactly within the right
range, for which such wrapping events occur and have been studied in
great detail, are the nucleoprotein capsids of many animal
viruses---belonging for instance to the families of Togaviridae,
Coronaviridae, Retroviridae, Rhabdoviridae, Or\-tho- and
Paramyxoviridae, and Hepadnaviridae \cite{GaHe98}.  During their final
maturation step the viral capsids are enveloped by a cellular membrane
(\eg\ the plasma membrane or the endoplasmic reticulum) in an event
which is believed to be independent of active cell processes and by
which they ultimately leave their host.  In the simplest case adhesion
is due to a direct interaction between the capsid and the membrane
(for instance in the case of type D retroviruses \cite{GaHe98}).
However, more common is an adhesion mediated by viral transmembrane
proteins (usually called ``spikes'') which can attach at specific
binding sites on the capsid \cite{GaSi74,SiGaHe82,LuKi00}, and for
which Semliki Forest Virus (SFV) is the classical example.  SFV has a
capsid radius of about $30\un{nm}$ and 80 spikes.  Assuming a typical
membrane bending stiffness of $\kappa\approx 20\,k_\romB T$, one finds
that the wrapping boundary ``W'' at $\tw=4$ corresponds to a binding
(free) energy per spike of about $6\,k_\romB T$, which is physically
reasonable.

The above estimates indicate that viral wrapping events can be
expected to take place in the low tension regime of the phase diagram,
close to the two phase boundaries.  This pattern is found time and
again in biology: Systems often seem to have evolved to lie close to
phase boundaries, because this permits large ``effects'' to be
triggered by comparatively small parameter changes.  Recall, however,
that the envelopment transition has been found to be discontinuous and
associated with a substantial energy barrier, which nature somehow has
to overcome.  A conceivable solution of this problem would be provided
by a coupling between curvature and compositional degrees of freedom
\cite{Lei86,KaAn93TaKa94,Sei93,JuLi96} with the result of enhancing
the concentration of lipid species in the highly curved rim which
actually prefers a high curvature.  This would lower the rim energy
and thereby the wrapping barrier.

It is crucial to realize that it is biologically feasible to actually
\emph{move} in the phase diagram of Fig.~\ref{fig:pd}.  For instance,
cells actively control and adjust their surface tension for the
purpose of surface area regulation \cite{MoHo01}.  Even more dramatic
changes in tension can occur when one switches between adhering
membranes.  If viral capsids get spontaneously wrapped, they evidently
must be in a region of the phase diagram in which the wrapped state is
stable (and, moreover, in which it is not rendered inaccessible by a
large barrier).  But the virus cannot stay wrapped forever.  As it
infects a new host cell, it typically becomes internalized via
receptor mediated endocytosis and ends up in an endosome, which it
again has to leave in order to avoid ultimately being digested by
cellular toxins.  Many viruses leave the endosome by fusing their
outer envelope with the endosomal bilayer, but if the capsid were too
strongly attached to the membrane, it could not be freed this way.  It
is usually assumed that the biochemical changes within the endosome
which lead to the fusion event (in particular, a lowering of pH) also
diminish the strength of adhesion.  However, within the theoretical
framework established in this paper it is tempting to speculate about
an alternative mechanism: If the bilayer tension of the endosome is
larger than the tension of the membrane at which the capsid became
enveloped, unwrapping can be efficiently promoted by vertically
crossing the phase boundary ``E'' from enveloped to partially wrapped,
as can be seen in Fig.~\ref{fig:pd}.  Moreover, the horizontal
adhesion axis of the phase diagram can not only be changed by
chemically modifying the spikes, but also by controlling their
\emph{density} in the membrane \cite{TzDe}.  This may not only
be relevant in the initial wrapping event, in which an increasing
density of spikes in the membrane can push the system over the
envelopment boundary, but also in the unwrapping process, when after
fusion the spikes can readily diffuse into the essentially spike-free
endosomal membrane and thus reduce the binding free energy.

%\vspace*{0.2em}

The above example illustrates how the physical principles discussed in
this paper can be directly relevant in a biological context.
Unfortunately it is often hard to disentangle them from other
biological processes or secondary effects of the experimental set-up.
Hence, a quantitative test of the present work appears more
practicable in well controlled model systems, \eg, similar to the ones
studied in Refs.~\cite{DiAn97,KoRa99}.  Nevertheless, the physical
results presented here can provide valuable insight into biological
problems which may complement other approaches.  As an example one
might think of a way to measuring cellular tension which is an
alternative to the current method of pulling a tether
\cite{HoSh96,DeJu02}.  The above analysis has shown how the degree of
wrapping of a colloid depends on the applied tension---in the regime
accessible by the small gradient expansion it is simply inversely
proportional, see Eqn.~(\ref{eq:z_sg}).  One can thus use suitably
coated colloids as \emph{tension probes}.  Unlike the tether approach
this method is in principle also applicable to \emph{intracellular}
membranes, even though a noninvasive determination of the degree of
wrapping will be very difficult for small beads.  The theory developed
in this work should then be useful for analyzing the results of such
measurements.

%%%%%%%%%%%%%%%%%%%%%%%%%%%%%%%%%%%%%%%%%%%%%%%%%%%%%%%%%%%%%%%%%%%%%%%%%%%
%%%%%%%%%%%%%%%%%%%%%%%%%%%%%%%%%%%%%%%%%%%%%%%%%%%%%%%%%%%%%%%%%%%%%%%%%%%

\begin{acknowledgments}

It is my pleasure to thank T. Bickel, W. M. Gelbart, S. Tzlil, and
A. Ben-Shaul for many stimulating discussions on the subject.
Financial support by the German Science Foundation under grant
De775/1-1 is also gratefully acknowledged.

\end{acknowledgments}

%%%%%%%%%%%%%%%%%%%%%%%%%%%%%%%%%%%%%%%%%%%%%%%%%%%%%%%%%%%%%%%%%%%%%%%%%%%
%%%%%%%%%%%%%%%%%%%%%%%%%%%%%%%%%%%%%%%%%%%%%%%%%%%%%%%%%%%%%%%%%%%%%%%%%%%

%%%%%%%%%%%%%%%%%%%%%%%%%%%%%%%%%%%%%%%%%%%%%%%%%%%%%%%%%%%%%%%%%%%%%%%%%%%


\begin{thebibliography}{99}
\bibitem{GaussBonnet}
  The product of the two principal curvatures is called the ``Gaussian
  curvature''.  Its integral over a surface $S$ can be rewritten as a
  line integral of the geodesic curvature over the boundary $\partial
  S$ of $S$ (Gauss-Bonnet-theorem \cite{Krey91}).  It is thus
  invariant under deformations of $S$ which change neither its
  boundary nor its topology.  Under these conditions the Gaussian part
  in Eqn.~(\ref{eq:Helfrich}) only results in an additive constant in
  the energy which can be ignored in what follows.
\bibitem{Lipowsky_barrier}
  In Ref.~\cite{LiDo98} it is suggested that for partially wrapped
  spheres the deformation of the membrane along the contact line will
  contribute an additional bending term to the contact line energy
  which acts as an energy barrier for the encapsulation process and
  which should scale $\sim\sqrt{z(2-z)}$.  However, if the membrane is
  tensionless, such a term is rigorously absent (due to its catenoid
  shape).  Under nonvanishing tension there is indeed a barrier, but
  it does not have the proposed form, since the bending energy is
  generally not localized at the rim.
\bibitem{catenoid_diverges}
  Observe that the stronger requirement $h(s)\rightarrow\text{const.}$
  is \emph{not} used.  In fact, for zero tension the solution of the
  shape equations is the catenoid, which \emph{diverges}
  logarithmically in the limit $s\rightarrow\infty$.  However, this is
  a special case, and for any nonzero tension the profile height
  $h(s)$ remains bounded and converges toward a constant.
\bibitem{bifurcate_relevance}
  In several cases of viral budding the nucleoprotein capsid is not
  preformed but aggregates simultaneously with the budding process
  \cite{GaHe98}.  It then imposes a similar boundary condition for the
  adhering membrane as a complete capsid, except that the membrane is
  not prevented from having a contact curvature larger than $1/a$
  (there is no capsid yet to exclude this region of space from being
  occupied).  In this case the states with $\dot\psi_0>1/a$ are no
  longer inaccessible and may become physically relevant.
\bibitem{catenoid_lambda}
  Note that as the neck contracts, the membrane deformation will
  ultimately occur on length scales small compared to the typical size
  $\lambda$ introduced in Eqn.~(\ref{eq:lambda}), hence tension ceases
  to be important.  But if only bending counts, the optimal solution
  is the catenoid, which has zero bending energy.  Indeed, in the
  limit $z\rightarrow 2$ the meridinal contact curvature $\dot\psi_0$
  converges toward $-1/a$, \ie, the negative of the azimuthal contact
  curvature.

  Of course, once the deformation of the membrane occurs on length
  scales comparable to its thickness (typically a few nanometers),
  neither linear elasticity theory underlying Eqn.~(\ref{eq:Helfrich})
  nor the description of the membrane as an idealized surface are
  appropriate, and the bilayer structure needs to be modeled in more
  detail.
\bibitem{fusion_speculation}
  Note also that the negative slope of $\tE(z)$ at $z=2$ (see inset of
  Fig.~(\ref{fig:barrier}) implies that the regions of the bilayer at
  the upper neck are pushed against each other.  This may potentially
  facilitate a subsequent fusion event.
\bibitem{LambertW}
  The (essentially) logarithmic divergence is obvious by looking at
  the inverse of $\CALW_{-1}(x)$, namely the function $x\rome^x$, which
  approaches 0 (essentially) exponentially from the negative side as
  $x$ approaches $-\infty$.  A more precise statement can be obtained
  by making use of the functional relation $\CALW_{-1}(x) = \ln|x| -
  \ln\big|\CALW_{-1}(x)\big|$, which can be iterated to get increasingly
  accurate expressions for $\CALW_{-1}(x)$.  Since 
  $\lim_{x\rightarrow0^-}\CALW_{-1}(x)/\ln|x|=1$, the lowest order reads
  $\CALW_{-1}(x) \; \approx \; \ln|x|$, and the next correction is then given by
  $\CALW_{-1}(x) \; \approx \; \ln|x| - \ln\big|\ln|x|\big|$.
\bibitem{psidot0}
  One more bit of support stems from Eqn.~(\ref{eq:pr0}), which
  suggests that $a\dot\psi_0$ and $\sqrt{1+2\ts z}$ are of the same
  order.  In fact, for $z=1$ one rigorously has
  $a\dot\psi_0=\sqrt{1+2\ts}$, for otherwise $p_r(0)$ would be
  infinite.
\bibitem{low_tension_exponent}
  It ought to be noted that in the absence of a theoretical
  description of the \emph{barrier height} in the low tension regime
  (a small gradient expansion is not enough), the exponent 0.86 is a
  purely empirical description of the numerical results, which are
  equally compatible with the form $a_1 \ts\times\log(a_2\ts)$.
\end{thebibliography}
\end{document}